\documentclass{iau}

\usepackage{amsmath}
\usepackage{graphicx}
\usepackage{multirow}

\begin{document}

\lefttitle{Ruth A. Daly}
\righttitle{Spin Angular Momentum and Black Hole Mass Components of Sgr A*}

\jnlPage{1}{6}
\jnlDoiYr{2026}
\doival{10.1017/xxxx}

\aopheadtitle{Proceedings IAU Symposium}
\editors{M. Zaja\v{c}ek,  T. Je\v{r}\'{a}bkov\'{a}, V. Karas, R. Schödel \&  P. Sukov\'{a}, eds.}
\bigskip
\title{The Spin Angular Momentum and Black Hole Mass Components of Sagittarius A*}

\author{Ruth A. Daly}
\affiliation{Department of Physics,  Pennsylvania State University, 
Berks Campus, Wyomissing, PA 19610, USA}

\begin{abstract}
The dimensionless spin angular momentum, dimensionless spin function,  
and black hole mass components of Sagittarius A* (Sgr A*) were obtained by Daly et al. (2024) by applying the outflow method to six independent sets of 
simultaneously or contemporaneously obtained X-ray and radio data. 
Consistent results were obtained with each data set. 
Several data set combinations were considered, 
and each yielded similar and consistent results. 
Set I  
was identified as the ``preferred" data set. 
Data Set I indicates that Sgr A* has a 
dimensionless 
spin angular momentum $\rm{a_*} = 0.90 \pm 0.06$ and 
a dimensionless spin function $\rm{F = 0.62 \pm 0.10}$.  
The results are consistent with the value of $\rm{a_*} = 0.93 \pm 0.15$ obtained with the outflow method applied to 
an independent data set (Daly 2019). 
Spin values obtained with the outflow method are independent of the orientation of the black hole spin axis relative to the observer and relative to the plane of the Galaxy. 
The empirically determined value of the dimensionless 
spin function, F, 
was applied to determine the rotational mass component, $\rm{M_{rot}}$, irreducible mass component, $\rm{M_{irr}}$, and the spin mass-energy available to be extracted from the black hole, $\rm{M_{spin}}$, 
relative to the total black hole mass, $\rm{M_{dyn}}$ using  
the expressions derived by Daly (2022).  
Weighted mean values of $\rm{(M_{rot}/M_{dyn})} = 0.53 \pm 0.06$; $\rm{(M_{irr}/M_{dyn})} = 0.85 \pm 0.04$; and $\rm{(M_{spin}/M_{dyn})} = 0.15 \pm 0.04$ were obtained for Sgr A* with the ``preferred" data set. 
Combining these with an  independent measure of the total black hole mass, $\rm{M_{dyn}}$, indicated values of $\rm{M_{rot} = (2.2 \pm 0.3) \times 10^6~ M_{\odot}}$; $\rm{M_{irr} = (3.5 \pm 0.2) \times 10^6~ M_{\odot}}$; and $\rm{M_{spin} = (6.2 \pm 1.6) \times 10^5 ~M_{\odot}}$ for Sgr A*.  
The black hole spin function, F, provides both a direct measure of the ratio $(\rm{M_{rot}/M_{irr}})$, and  
of the angular velocity of the hole 
at the black hole horizon, $\rm{\Omega_H}$, relative to the value 
expected for a maximally spinning black hole, $\rm{\Omega_{H,max},~ F = (\Omega_H/\Omega_{H,max})}$. 
The spin mass-energy available to be extracted from the hole, $\rm{M_{spin}}$, is a particularly important quantity since the extraction of this mass-energy 
can significantly impact the source environment, and processes that extract the spin energy 
necessarily increase the black hole irreducible mass. 
The application of the outflow method 
to a weak compact radio 
source such as Sgr A* in its current state is most accurately determined with simultaneous or contemporaneous radio and X-ray measurements, though these are not necessarily expected to be produced co-spatially. Here, tests of the outflow method are discussed, and results obtained for Sgr A* with the outflow method are reviewed. 
\end{abstract}

\maketitle
\section{Introduction}
The two parameters that quantify the properties of an astrophysical black hole are the total black hole mass, $\rm{M_{dyn}}$, and the  
dimensionless spin angular momentum, $\rm{a_*}$, or the dimensionless spin function, F, 
of the hole.
Combining $\rm{a_*}$, or the spin function F, 
and $\rm{M_{dyn}}$, allows the determination of three key black hole mass, or mass-energy, components: the irreducible mass, 
$\rm{M_{irr}}$, the rotational mass-energy, 
$\rm{M_{rot}}$, and the spin mass-energy available to be extracted 
from the hole, 
$\rm{M_{spin}}$, 
where $\rm{M_{dyn}^2 = M_{rot}^2 + M_{irr}^2}$ 
and 
$\rm{M_{spin} = M_{dyn} - M_{irr}}$ (Penrose 1969; Christodoulou 1970; Christodoulou \& Ruffini 1971; 
Hawking 1971; Penrose \& Floyd 1971; Thorne et al. 1986; Blandford 1990; Daly 2022).  
The dynamical mass, $\rm{M_{dyn}}$, is the mass that will be 
measured by a distant observer using standard astronomical techniques. 
The total black hole mass, $\rm{M_{dyn}}$, of Sgr A* has been studied for some time, and there is now a consensus on a total black hole mass value of 
about $4.297 \times 10^6 M_{\odot}$ 
(Gravity Collaboration 2023). However, it is only recently that  
the spin properties of Sgr A* have started to converge and perhaps 
are moving toward a 
consensus value, as discussed in section 5. 

One method to empirically determine the black hole 
 dimensionless spin angular momentum, 
$\rm{a_*}$, and dimensionless spin function, $\rm{F}$, is the outflow method (Daly 2019; Daly et al. 2024). The outflow method 
does not rely 
upon a specific accretion disk or jet-launching model  
and does not depend upon the orientation of 
the observer relative to the black hole spin axis. 
The method is described in 
detail by Daly (2019), and is summarized in section 3 of 
Daly et al. (2024). The expressions used to obtain black hole 
mass component ratios directly from the spin function, F, are 
presented by Daly (2022). 

Here, tests of the outflow method are described in sections 2 and 3.  
Results obtained for the mass components of Sgr A* are discussed in 
section 4. Values of the dimensionless spin angular 
momentum of Sgr A* obtained with the outflow method are presented   
and compared with those obtained independently 
in section 5. It should be noted that for all types of outflows studied 
to date, including both stellar-mass and supermassive black holes, 
the amount of spin mass-energy extracted per outflow event is a small 
fraction of the spin mass-energy available to be extracted, so the 
outflow does not significantly impact the dimensionless spin angular momentum of the black hole, and the black hole spin is expected to 
remain constant over human timescales (e.g. Kolo{\v s} et al. 2021; Stuchl\'{i}k et al. 2021;  
Rueda \& Ruffini 2023), as indicated empirically by results obtained with the outflow method (Daly 2019, 2022, 2025).

\section{Tests of the Outflow Method with Stellar Mass Black Holes}

For the stellar mass black holes GX 339-4, A0 6200, and V404 Cyg,   
which vary on human timescales, the data of Saikia et al. (2015) allowed the determination of numerous independent values of the black hole spin function, 
$\rm{F}$, dimensionless spin angular momentum, 
$\rm{a_*}$, and other quantities 
(see Table 2 of Daly 2019). 
The quantities listed in Table 2 were combined to obtain mean values and standard deviations 
for each of the stellar-mass black holes mentioned above (see Table 1 of Daly 2019).  In that work, different notation was used and 
the luminosity in directed kinetic energy, 
$\rm{L_{dKE}}$ is referred to as 
$\rm{L_j}$; $\rm{F}$ is referred to as $\rm{\sqrt{f(j)/f_{max}}}$;    
and $\rm{a_*}$ is referred to as $\rm{j}$. 

GX 339-4, Test (i): Comparisons of $\rm{a_*}$ obtained with the outflow method and the X-ray reflection method, which does not include any data or 
information related to an outflow, 
are included in columns 5 and 6 of Table 1 of Daly (2019).  
Excellent agreement is obtained for the 
dimensionless spin angular momentum. The spin value obtained with the outflow method is 
$a_* = 0.92 \pm 0.06$, which is compared with those obtained independently with the X-ray reflection method of $a_* = 0.94 \pm 0.02$ (Miller et al. 2009) and 
 $a_* = 0.95^{+0.03}_{-0.05}$ (Garc\'{i}a et al. 2015). This indicates that the outflow method is providing reliable results since  
 there is no overlap between the application of the X-ray reflection and outflow methods.

GX 339-4, Test (ii): The fact that the standard deviation of Log(F) of 0.06, which is the quantity listed in column 4 of Table 1 of Daly (2019),  
is 3 times smaller than that of $\rm{Log(B/B_{Edd})}$ of 0.18, 
listed in column 7 of Table 1, 
indicates that the outflow method is providing reliable results. 
That is, even 
though the values of quantities listed in Table 2, which 
were combined to obtain the mean values and standard 
deviations listed in Table 1, 
vary with time leading to the rather large standard deviation of $\rm{Log(B/B_{Edd})}$, 
where $\rm{(B/B_{Edd})}$ is the magnetic field strength in 
Eddington units (see the definitions in Daly 2019), 
the combination of the bolometric luminosity, 
$\rm{L_{bol}}$, and luminosity in directed kinetic energy, 
$\rm{L_{dKE}}$, vary in such a way as to keep the spin 
function, F, relatively constant.  

A0 6200, Test (i): Data for 
A0 6200 allows a comparison that provides particularly strong support for the outflow method. This arises from the comparison of results obtained with the outflow method applied to data obtained when the source was in a relatively quiescent phase,  
with those obtained 
when the source was in outburst and was about a million times more luminous, as described below. 
The first set of values for this source, 
obtained with data published by Saikia et al. (2015) when 
the source was in a quiescent phase, are 
listed in Tables 1 and 2 of Daly (2019) and 
indicate a value of $\rm{a_* = 0.98 \pm 0.07}$. 
A second set of measurements for this source, 
determined with radio and X-ray data obtained contemporaneously in 1975 when the source was undergoing an X-ray outburst during which 
$\rm{L_{bol}}$ increased by about 6 orders of magnitude and 
$\rm{L_{dKE}}$ increased by about 3 orders of magnitude relative to the values obtained with the 
Saikia et al. (2015) data, are
discussed in section 5 of Daly (2019). 
When evaluated in the context of the outflow method, the 1975 data 
indicate a value of 
$\rm{a_* = 0.97 \pm 0.07}$. Thus, 
even though during the outburst 
$\rm{L_{bol}}$ increased by about 6 orders of magnitude and 
$\rm{L_{dKE}}$ increased by about 3 orders of magnitude relative to the values obtained with the 
Saikia et al. (2015) data, 
the outflow method indicated a very similar value 
for $a_*$. 
This suggests that the outflow method is providing consistent and reliable measures of black hole 
spin properties. 

V404 Cyg, Test (i): Table 1 of Daly (2019) indicates that the standard deviation  
of $\rm{Log(F)}$ is about 0.06, while that for $\rm{Log(B/B_{Edd})}$ is 4 times larger, and is about 0.24. As with GS 339-4, 
this indicates that $\rm{L_{dKE}}$ and $\rm{L_{bol}}$ each vary substantially, however, when combined to obtain $\rm{Log(F)}$ this quantity remains relatively stable and 
thus has a much smaller standard deviation than $\rm{Log(B/B_{Edd})}$. This is as expected in the outflow method, as described above for GS 339-4. 

\section{Tests of the Outflow Method with  Supermassive Black Holes}
Two types of tests can be applied to supermassive 
black holes. Sources that exhibit compact radio emission, 
and thus are similar to Sgr A*, which are listed in Table 3 of Daly (2019), are discussed in part (1) below. Powerful extended classical double radio sources, which are listed in Table 4 of Daly (2019),  
are discussed in part (2) below. 

(1) A comparison of spin values obtained with the outflow method and the X-ray reflection method was possible for six supermassive black holes, Ark 564, Mrk 335, NGC 1365, 
NGC 4051, NGC 4151, and 3C120, 
and these comparisons 
are listed in Table 1 of Daly (2019).  
These comparisons 
indicate reasonable agreement between spin 
values obtained in the contexts of the X-ray reflection method and the outflow method. 
This indicates that the outflow method is providing reliable results since  
 there is no overlap between the application of the X-ray reflection and outflow methods. 
The supermassive black holes included in 
Table 1 are 
of the same type as Sgr A* (see Table 3 of Daly 2019), and exhibit compact, relatively localized radio emission. Spin values for Sgr A* and M87* are also included in Table 1. 

(2) Black hole spin properties obtained with independent methods for  
supermassive black holes with very powerful, large-scale, 
collimated outflows that typically produce classical double radio sources  also provide empirical support for the outflow method (e.g. Daly 2019, 2022; Azadi et al. 2023).  
Dimensionless spin angular momentum values, $a_*$ and 
spin functions, F, for a sample of 100 very powerful 
classical double radio sources with redshifts between about zero and two 
were obtained and studied by Daly (2019, 2022); in that work $a_*$ 
is referred to as $j$.  
The sources are known as FRIIb sources and their properties are summarized by Daly (2025). 
The distribution of spin function values 
are studied by Daly (2022), who found that about 2/3 of the supermassive black holes are maximally spinning, while 1/3 of have a broad distribution extending to lower F values (see Fig. 5 of that work). 

A subset of about 15 of the 100 sources are included in the detailed source modeling study of Azadi et al. (2023), who considered a sample of 20 sources in the context of a continuum fitting method, which has no overlap 
with the outflow method. 
Azadi et al. (2023) conclude that there is good agreement between the 
independently determined spin values. 
This provides strong support for the outflow method since the outflow method 
is independent of any particular jet-launching model and any particular accretion disk model (e.g. Daly 2016, 2019, 2022, 2025), while the results of 
Azadi et al. (2023) are obtained in the context of detailed modeling of emission at all wavelengths from the sources. 

\section{Results Obtained for the Black Hole Mass Components of Sagittarius A*}
Daly et al. (2024) obtain, perhaps for the first time, measurements of the 
mass components of the supermassive black hole Sagittarius A*. The application of the outflow method provides a empirical determination of the black hole 
spin function, $\rm{F}$ (Daly 2019). 
Physically, the spin function $\rm{F}$ is equivalent to the ratio 
of the rotational mass component to the irreducible mass component 
of Sgr A* since 
$\rm{F = (M_{rot}/M_{irr})}$ (Daly 2022). In that work, it is shown that 
values of the spin function, $\rm{F}$, map directly to  
the rotational, irreducible, and spin mass components of the  
black hole 
relative to the total black hole mass, $\rm{M_{dyn}}$. The black hole spin function 
is also related to 
the angular velocity at the horizon of the black hole relative to the value expected for a maximally spinning black hole, $\rm{F = (\Omega_H/\Omega_{H,max}})$.  
Thus, the value of $\rm{F = 0.62 \pm 0.10}$ obtained 
by Daly et al. (2024) with the ``preferred" data set indicates that 
Sagittarius A* is rotating with an angular 
velocity that is $0.62 \pm 0.10$ of the 
maximum 
possible value. 

The relationships derived by Daly (2022) between 
the spin function $\rm{F}$ and black hole mass components 
were applied by Daly et al. (2024) to obtain 
values of the black hole mass components 
$\rm{M_{rot}, M_{irr}}$, and $\rm{M_{spin}}$ relative to the 
total dynamical black hole mass, $\rm{M_{dyn}}$.
Values of 
$\rm{(M_{rot}/M_{dyn})} = 0.53 \pm 0.06$; $\rm{(M_{irr}/M_{dyn})} = 0.85 \pm 0.04$; \&  $\rm{(M_{spin}/M_{dyn})} = 0.15 \pm 0.04$ are obtained for Sagittarius A* 
by Daly et al. (2024) 
with the ``preferred data set;" similar values are obtained with
individual data sets and other combinations of data sets 
(see Tables 5, 6, and 7 of that work). 

To solve for each black hole mass component, a total mass of 
$\rm{M_{dyn}} = (4.152 \pm 0.014) \times 10^6 \rm{M_{\odot}}$ (Gravity 
Collaboration 2019) was adopted by Daly et al. (2024).  
Values for $\rm{M_{rot}, M_{irr}}$, 
and $\rm{M_{spin}}$ obtained for each data set are 
listed in Table 5 of Daly et al. (2024); values for 
data set I, the ``preferred data set," are listed in Table 6 
and indicate values of 
$\rm{M_{rot} = (2.2 \pm 0.3) \times 10^6~ M_{\odot}}$; $\rm{M_{irr} = (3.5 \pm 0.2) \times 10^6~ M_{\odot}}$; and $\rm{M_{spin} = (6.2 \pm 1.6) \times 10^5 ~M_{\odot}}$ for Sgr A*. 
The results indicate that the spin mass-energy available to 
power an outflow from Sgr A*, $\rm{M_{spin}}$, is small in both relative and 
absolute terms: $\rm{(M_{spin}/M_{dyn})} = 0.15 \pm 0.04$ and 
$\rm{M_{spin} = (6.2 \pm 1.6) \times 10^5 ~M_{\odot}}$ for 
the preferred data set; similar results are obtained 
with other data sets. 
Thus, it is not surprising that Sgr A* is a weak radio source, as discussed in sections 5.2 and 5.3   of 
Daly et al. (2024).
Similar values for all of the quantities discussed here are obtained with other combinations of data sets, as summarized in Tables 6 and 7, and as discussed in section 5 of that work. 

\section{Results Obtained for the Dimensionless Spin Angular Momentum of Sgr A*}

The first dimensionless spin angular momentum value obtained for 
Sgr A* with the outflow method is $\rm{a_*} = 0.93 \pm 0.15$   
(Daly 2019) (see Table 1 of that work). 
Six new independent values are obtained with the outflow method 
by Daly et al. (2024); these data sets are labeled C1, C2, C3, C4, B1, and B4 (see Tables 1, 2, and 3 and the Notes following Table 6). The data sets labeled with a "C," that is, C1, C2, C3, and C4, 
include hundreds of radio observations, and thus are considered to be highly reliable.
Radio data sets C1 and C4 
are obtained simultaneously with Chandra X-ray data, while data sets C2 and C3 are obtained within one day of the Chandra data. It is noted that an X-ray flare event occurred during observing run C3, which is also contemporaneous with B3; and data set B2 is contemporaneous with C2. 
A literature search identified 2 additional single radio observations that occurred contemporaneously with 2 additional Chandra observations, labeled B1 and B4. 

Data sets C1, C2, and C4 are identified as ``preferred" since they include numerous radio observations and the flaring X-ray event was excluded. 
A value of $a_* = 0.90 \pm 0.06$ is obtained with this combined data 
set, as indicated in Table 6 of Daly et al. (2024), identified as ``Set I." 
Considering all of the 5 non-flaring Chandra data sets (that is,   
adding data sets B1 and B4 to Set I), led to Set II, which indicates 
a value of $a_* = 0.92 \pm 0.04$. Considering only data with numerous radio observations and including the X-ray flaring event, labeled Set III (which includes data sets C1, C2, C3, and C4), indicates a value of 
$a_* = 0.87 \pm 0.06$. And, including each Chandra data set once, labeled 
Set IV (which adds data sets B1 and B4 to Set III), indicates a value of 
$a_* = 0.90 \pm 0.04$. This is identical to the value obtained with Set I and has a slightly smaller uncertainty. 

Overall, results obtained with data set I, listed in Tables 6 and 7, of Daly et al. (2024) are taken to be the ``preferred" values. 
Thus, the preferred value obtained with the outflow method by Daly et al. (2024) 
is $\rm{a_*} = 0.90 \pm 0.06$.  This is consistent with the value 
of $0.93 \pm 0.15$ obtained 
with an independent data set by Daly (2019). 

Several recent values reported for Sgr A* are in broad agreement with these values. Eckart et al. (2018) provide an overview of published spin values obtained with several model-dependent radio, near-infrared, and X-ray data analyses, and from this list a slight tendency for $a_* > 0.5$ crystallizes.
Witzel et al. (2018) present data which are consistent with a value of $a_* = 0.92$, although no uncertainty is cited (see their Fig. 14).  
The Event Horizon Telescope 
(EHT) group select 5 values of $a_*$ of -0.94, -0.5, 0.0, 0.5, and 0.94, run numerous model-dependent simulations, and find that the best agreement is obtained for $a_* =0.94$, though the agreement is not perfect (EHT 2022).
Dokuchaev (2023) infers spins of $0.65 < a_* < 0.9$
for Sgr A* and $a > 0.75$ for M87*, which are consistent with the values obtained with the outflow method for Sgr A*, noted above,  and the value of $1.0 \pm 0.2$ obtained 
for M87* with the outflow method (Daly 2019). 
Yfantis et al. (2024) fit a hot-spot model to the light curves of Sgr A* 
and find that $a_* > 0.8$. 
Andrianov \& Chernov (2024) obtain an estimate of 
$a_* \approx 0.9$ by modeling spots seen by EHT as flares near the black hole horizon. 
Janssen et al. (2025) obtain 
a value of $a_* \sim (0.8 - 0.9)$ for Sgr A* by applying detailed modeling of EHT data in the context of the Zingularity framework. 
Overall, it appears that evidence is mounting that Sagittarius A* has 
a spin value that is consistent with $a_* = 0.93 \pm 0.15$ and 
$a_* = 0.90 \pm 0.06$ obtained with independent data sets in the context of  the outflow method. 

{\bf{Acknowledgments}}
It is a pleasure to thank my collaborators in these endeavors, especially Megan Donahue, ~Daryl Haggard, ~Anan Lu, ~Chris O'Dea, and ~Biny Sebastian.


\begin{thebibliography}{}
\bibitem[Azadi et al. (2023)]{Azadi2023}
Azadi, M., Wilkes, B., Kuraszkiewicz, J., McDowell, J., Siebenmorgen, R., Ashby, M., Birkinshaw, M., Worrall, D., Abrams, N., Barthel, P., Fazio, G. G., Haas, M., Hyman, S., Martínez-Galarza, R., \& Meyer, E. T. 2023, ApJ 945, 145
\bibitem[Andrianov (2024)]{AC24}
Andrianov, A. S., \& Chernov, S. V. 2024, Astronomy Reports, Vol. 68, 233  
\bibitem[Blandford (1990)]{Blandford1990} 
Blanford, R. D. 1990, in Active Galactic Nuclei, eds. 
T. J.-L. Courvoisier \& M. Mayor (Springer-Verlag, Berlin Heidelberg), (161 - 269). 
\bibitem[Christodoulou(1970)]{1970PhRvL..25.1596C} Christodoulou, D.\ 1970, Phys. Rev. Lett., 25, 1596
\bibitem[Christodoulou \& Ruffini (1971)]{CR71} Christodoulou, D. Ruffini, R. Reversible Transformations of a Charged Black Hole.
{Phys. Rev. D}, 
{1971}, 3552. 
\bibitem[Daly(2019)]{Daly2019}
Daly, R. A.\ 2019, ApJ, 886, 37
\bibitem[Daly (2022)]{D2022} 
Daly, R. A. 2022, MNRAS, 517, 5144
\bibitem[Daly(2025)]{Daly 2025} Daly, R. A. 2025, 
Universe, 11(8), 267
\bibitem[Daly et al. (2024)]{Daly et al. 2024}
Daly, R. A., Donahue, M., O'Dea, C. P., Sebastian, B., Haggard, D., \& Lu, A. 2024, MNRAS, 527, 428
\bibitem[Dokuchaev (2023)]{Dokuchaev23}
Dokuchaev, V. I. 2023, Astronomy, 2, 141
\bibitem[Eckart et al. (2018)]{Eckart2018}
Eckart, A., Tursunov, A. A.,  Zaja\v{c}ek, M., Parsa1, M., Hossein, 
E., Subroweit, M., Peissker, F., Straubmeier, C., Horrobin1, M.,  Karas, V. 2018, ACPS, 342, 48
\bibitem[EHT (2022)]{EHT22}
Event Horizon Telescope Collaboration (EHT) 2022, ApJL, 930, L12  
\bibitem[Garcia et al. (2015)]{Garcia2015}
Garc\'{i}a, J. A., Steiner, J. F., McClintock, J. E., Remillard,
R. A., Grinberg, V., \& Dauser, T. 2015, ApJ, 813, 84
\bibitem[GC (2019)]{GC2019} 
GRAVITY Collaboration 2019, A\&A, 625, L10 
\bibitem[GC (2023)]{GC2023} 
GRAVITY Collaboration, 2023, A\&A, 677, L10 
\bibitem[Hawking (1971)]{Hawking1971}Hawking, S.W. 1971, PhysRevLett, 26, 1344
\bibitem[Janssen et al. (2025)]{Janssen25}
Janssen, M., Chan, C. K., Davelaar, J. \& Wielgus, M 
2025, A\&A, 698, A62
\bibitem[Kolo\v s et al. (2021)]{Ketal21}
Kolo\v s M., Tursunov, A., \& Stuchl\'ik, Z. 2021, 
{Phys. Rev. D}, {103}, 24021
\bibitem[Miller et al. (2009)]{Milleretal2009}
Miller, J. M., Reynolds, C. S., Fabian, A. C., Miniutti, G.,
\& Gallo, L. C. 2009, ApJ, 697, 900
\bibitem[Penrose (1969)]{P1969} Penrose, R. 1969, Riv. Nuovo Cim., 1, 252
\bibitem[Penrose \& Floyd (1971)]{PF1971} Penrose, R., \& Floyd, R. M. 
1971, NPhS, 229, 177
\bibitem[Rueda \& Ruffini (2023)]{RR23}
{Rueda, J.A., \& Ruffini, R.} 2023, {Eur. Phys. J. C} {83}, 960
\bibitem[Saikia et al. (2015)]{Saikiaetal2015}
Saikia P., K\"{o}rding E., \& Falcke H., 2015, MNRAS, 450,
2317
\bibitem[Stuchl{'}ik et al. (2021)]{Setal21}
Stuchl\'ik, Z., Kolo\v s, M., \& Tursunov, A. 2021, 
{Universe}, {2021}, {7}, 416 
\bibitem[Thorne et al. (1986)]{T86}
Thorne, K. S., Price, R. H., Macdonald, D. A., Wai-Mo, S., 
\& Zhang, X 1986, in "Black Holes: The Membrane Paradigm," (67-120)
\bibitem[Witzel et al. (2018)]{W18}
Witzel, G., Martinez, G., Hora, J., Willner, S. P., Morris, M. R., Gammie, C.,  Becklin, E. E., Ashby, M. L. N.,  Baganoff, F., Carey, S., Do, T., Fazio, G. G., Ghez, A., Glaccum, W. J., Haggard, D., Herrero-Illana, R., Ingalls, J., Narayan, R., \& Smith, H. A.
2018, ApJ, 863, 15
\bibitem[Yfantis et al. (2024)]{Yfantis24}
Yfantis, A. I., Mo\'scibrodzka, M. A., Wielgus, M., Vos, J. T., \& 
Jimenez-Rosales, A. 2024, A\&A, 685, A142
\end{thebibliography}
\end{document}